\documentclass[prl,aps,showkeys,showpacs,superscriptaddress,twocolumn,%
amsmath,amssymb]{revtex4}
\usepackage{dsfont,graphicx}

\newcommand{\grad}{\nabla}
\newcommand{\bfB}{{\mathbf B}}
\newcommand{\bfE}{{\mathbf E}}
\newcommand{\bfJ}{{\mathbf J}}
\newcommand{\bfL}{{\mathbf L}}
\newcommand{\bfP}{{\mathbf P}}

\newcommand{\bfS}{{\mathbf S}}

\newcommand{\bfk}{{\mathbf k}}

\newcommand{\bfr}{{\mathbf r}}

\newcommand{\bfx}{{\mathbf x}}

\begin{document}

\title{Detecting orbital angular momentum in radio signals}

\author{H.\,Then}
\affiliation{%
  Institute of Physics, 
  Carl-von-Ossietzky Universit\"at Oldenburg,
  D-261\,11 Oldenburg,
  Germany}%

\author{B.\,Thid\'e}
\affiliation{%
  Swedish Institute of Space Physics,
  \AA ngstr\"om Laboratory,
  P.\,O.\,Box 537,
  SE-751\,21 Uppsala,
  Sweden}%
\affiliation{%
  LOIS Space Centre, V\"axj\"o University,
  SE-351\,95 V\"axj\"o, Sweden}%

\author{J.\,T.\,Mendon\c{c}a}
\affiliation{%
  CFP and CFIF, Instituto Superior T\'ecnico, PT-1096 Lisboa, Portugal}%

\author{T.\,D.\,Carozzi}
\affiliation{%
  Astronomy and Astrophysics Group,
  Department of Physics and Astronomy,
  University of Glasgow,
  Glasgow, G12\,8QQ,
  Scotland,
  United Kingdom}%

\author{J.\,Bergman}
\affiliation{%
  Swedish Institute of Space Physics,
  \AA ngstr\"om Laboratory,
  P.\,O.\,Box 537,
  SE-751\,21 Uppsala,
  Sweden}%

\author{W.\,A.\,Baan}
\affiliation{%
  ASTRON, P.\,O.\,Box 2, NL-7990 AA Dwingeloo, Netherlands}%

\author{S.\,Mohammadi}
\affiliation{%
  Department of Astronomy and Space Physics,
  Uppsala University,
  P.\,O.\,Box 515,
  SE-751\,20 Uppsala,
  Sweden}%

\author{B.\,Eliasson}
\affiliation{%
  Theoretische Physik IV,
  Ruhr-Universit\"at Bochum,
  D-447\,80 Bochum,
  Germany}%
\affiliation{%
  Department of Physics,
  Ume{\aa} University,
  SE-901\,87 Ume{\aa},
  Sweden}%

\begin{abstract}
Electromagnetic waves with an azimuthal phase shift are known to have
a well defined orbital angular momentum. Different methods
that allow for the detection of the angular momentum are proposed.
For some, we discuss the required
experimental setup and explore the range of applicability.
\end{abstract}


\maketitle

\section{Introduction}

It is known from Maxwell's theory that electromagnetic (em) radiation
carries angular momentum \cite{Abraham1914} which can typically be
separated into spin and orbital angular momentum associated with
polarization and spatial distribution, respectively. The photon spin
angular momentum (PSAM) has routinely been used for several decades;
and the progress in optical studies of photon orbital angular
momentum (POAM) has been rapid since Heckenberg et al.
\cite{Heckenberg1992b} pointed out that laser modes with well-defined
POAM can be readily produced and be detected via holograms. The POAM
has been extensively used in atomic and molecular physics
\cite{CohenTannoudji1998}. It has also been employed in quantum optical
communication concepts \cite{Gibson2004}, for entangling photons
\cite{Mair2001}, and for manipulating and orienting small particles
trapped in optical tweezers \cite{Ladavac2004}. Meanwhile it has
become impossible to list all applications; for reviews see
\cite{Allen1999,Allen2002,Santamato2004}.

All reported
experiments have in common that they superimpose phase singularities
(vortices) in optical fields. In a plane perpendicular to the beam
axis, the phases of the electric and magnetic vector fields have an
$l\phi$ dependence where $l$ is an integer and $\phi$ is the azimuthal
angle. This means that for $l\not=0$ the phase fronts of beams are not
planar but helical. As shown in Ref. \cite{Allen1992}, this implies
that the beam carries an OAM of $l\hbar$ per photon; see also Ref.
\cite{Harwit2003}. Because of destructive interference, the vortices
appear as isolated dark spots and give rise to doughnut shaped beam
profiles.

However, besides the enormous advantages that POAM offers, it has apparently
never been used in the radio domain. Moreover, it was an open question
until recently of how to emit, manipulate, and detect POAM of radio waves.
All the methods that have successfully been applied in optics are
practicably unfeasible in radio. For instance, scaling a hologram that
fits within a few square millimeters for optical wavelengths,
$\lambda<1\,\mu$m, to one that is designed for radio waves, $\lambda>1\,$m,
results in a hologram that covers several square kilometers in size.

The goal of the present paper is to highlight a number of
possibilities that allow for the detection of POAM in radio.
For some methods we discuss the required experimental setup and
explore the range of applicability.

\section{Antenna arrays}

Antenna arrays are capable of emitting radio beams with a well defined
POAM as shown numerically in \cite{Thide2007} and analytically in
\cite{Then2008}. An advantage of radio is that the phase of the em wave
is directly accessible and that digital techniques are available for
low frequencies ($\le1\,$GHz) which allow to manipulate the POAM in
software.

\begin{figure}[t]
  \begin{minipage}{\columnwidth}
    \includegraphics[width=0.4\columnwidth]{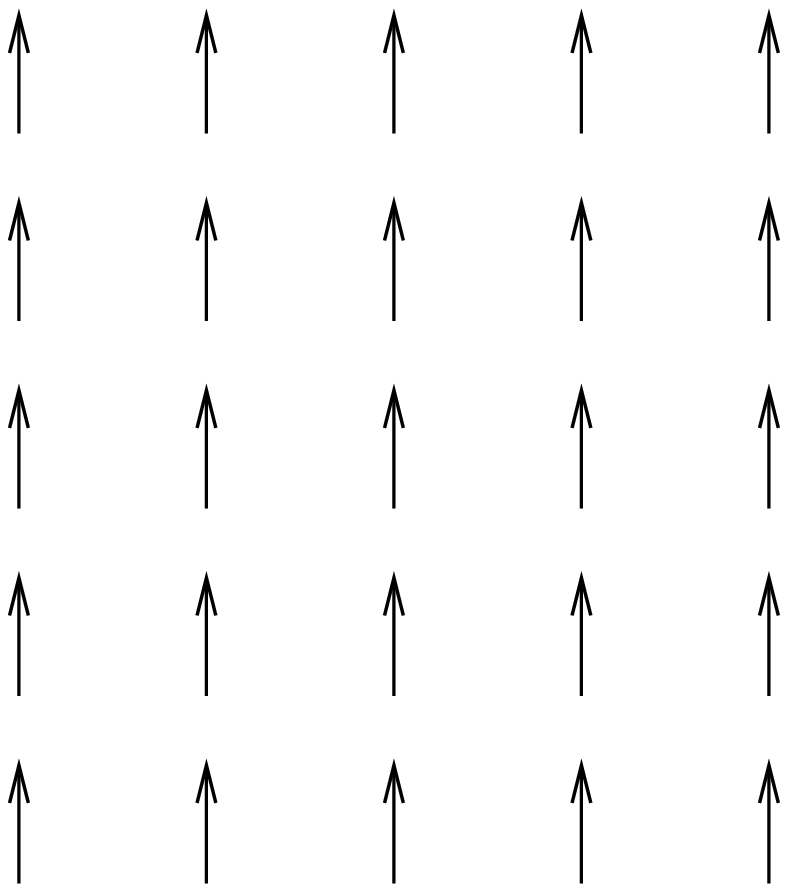}
    \hfill
    \includegraphics[width=0.4\columnwidth]{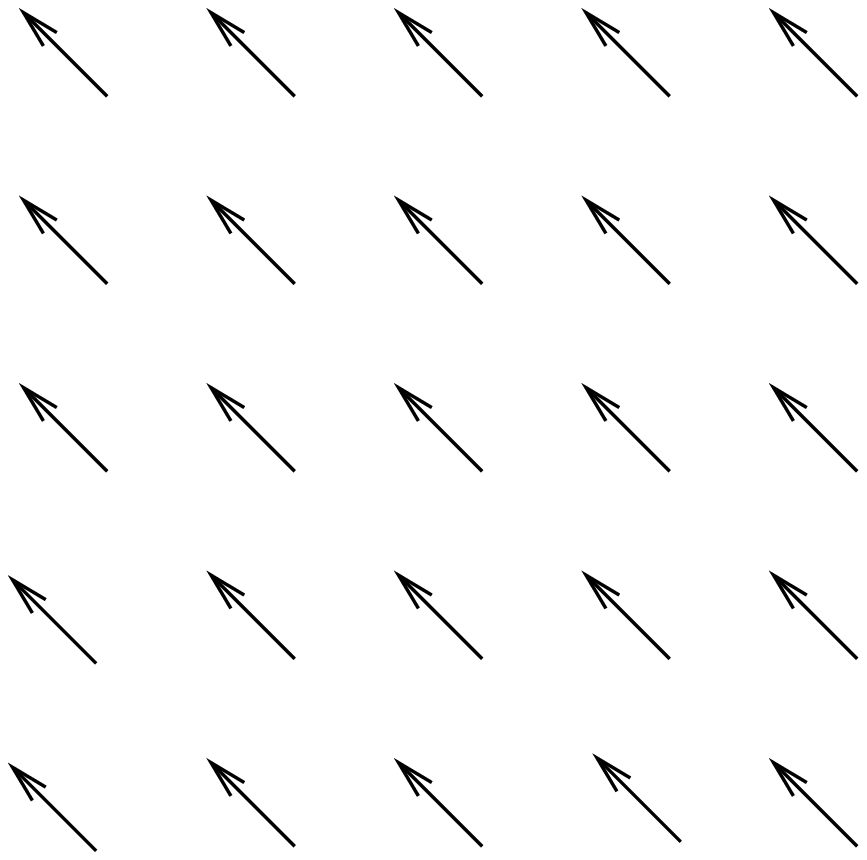}
  \end{minipage}
  \caption{%
    Schematic description of the instantaneous direction of the
    electric field vector of a radio beam with spin angular momentum
    $\bfS$ across a $5\times5$ antenna array. The left panel is for
    $\omega t=0$, and the right panel is for $\omega t=\frac{\pi}{4}$.}
  \label{fig:1}
\end{figure}

\begin{figure}[t]
  \begin{minipage}{\columnwidth}
    \includegraphics[width=0.4\columnwidth]{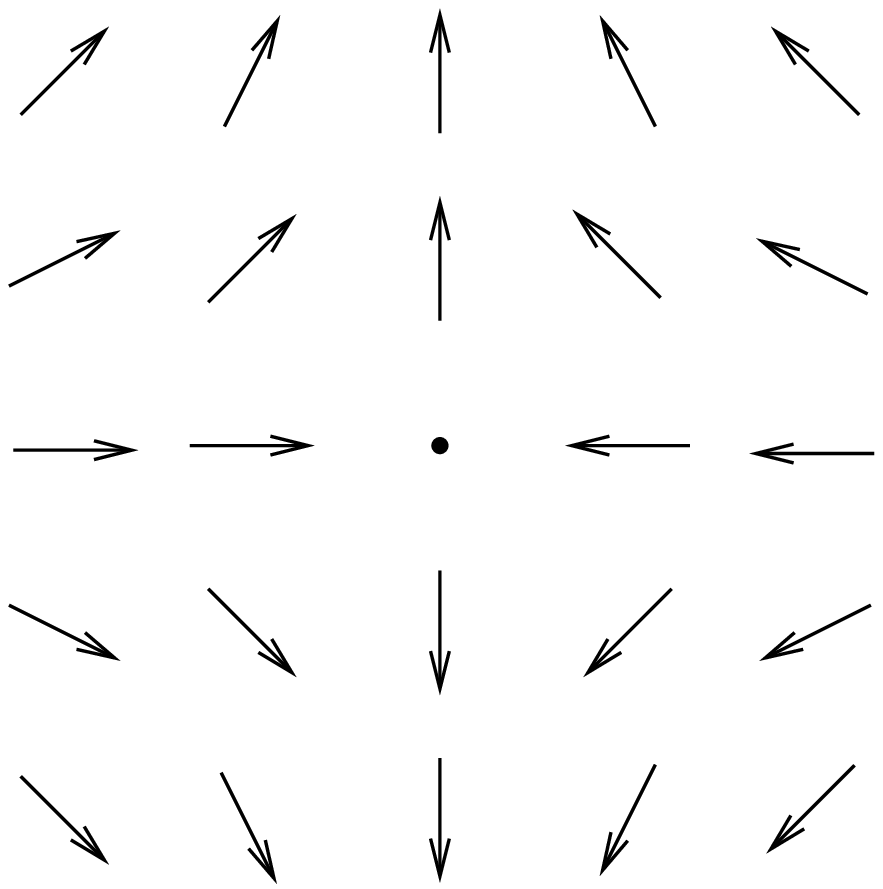}
    \hfill
    \includegraphics[width=0.4\columnwidth]{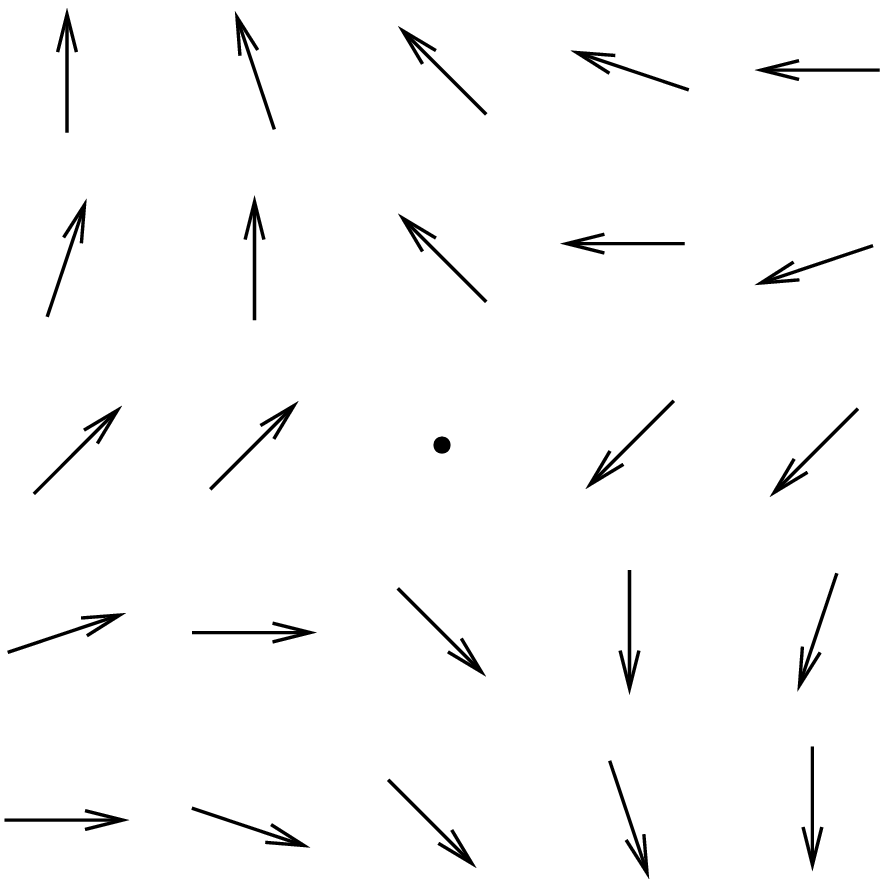}
  \end{minipage}
  \caption{%
    Schematic description of the instantaneous direction of the
    electric field vector of a radio beam with spin and orbital angular
    momentum, $\bfS$ and $\bfL$, across a $5\times5$ antenna array.
    The left panel is for $\omega t=0$, and the right panel is for
    $\omega t=\frac{\pi}{4}$.}
  \label{fig:2}
\end{figure}

For a pure spin state, the
handedness of the circular polarization is the same at any point in
the upper half space, and hence, can be measured with a single
tripole antenna. But the POAM is non-local and in order to detect
it, one needs full knowledge of the azimuthal phase shift around the
invariant beam axis. Figure \ref{fig:1} displays the instantaneous
directions of the electric field vector of a radio wave with spin
angular momentum $\bfS$ across a $5\times5$ antenna array, and
Fig. \ref{fig:2} displays the instantaneous directions of a radio
wave with spin and orbital angular momentum, $\bfS$ and $\bfL$.
Since the displayed beam is right circular polarized,
$\bfE(\bfr,t)=\frac{1}{\sqrt{2}}(1,i,0)E(\bfr)e^{-i\omega t}$,
the direction of the instantaneous electric field vector can be
identified with the local instantaneous phase of the em wave.
In order to measure the POAM of the em wave that is
displayed in Fig. \ref{fig:2} one needs at least three antennas
that are phase correlated with each other \cite{Thide2005a,Thide2005b}.

\section{Direct torque measurement}

\begin{figure}[t]
  \begin{minipage}{\columnwidth}
    \includegraphics[width=\columnwidth]{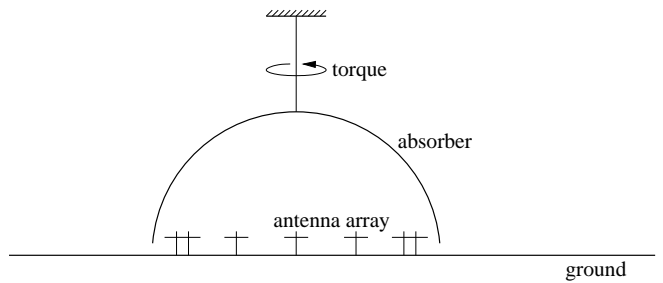}
  \end{minipage}
  \caption{%
    A radio experiment that measures the torque of the
    total angular momentum flux.}
  \label{fig:3}
\end{figure}

The torque of the angular momentum flux can be measured with an
absorber, e.g. a half sphere of graphite, that surrounds the antenna
array, see Fig. \ref{fig:3}.
If the antenna array is directly over ground,
the torque is
$
  \frac{dJ_z}{dt}=\alpha\frac{l+h}{\omega}\frac{dW}{dt},
$
where $\frac{dW}{dt}$ is the power that is radiated by the
antenna array and $\alpha$ is a constant depending on  the reflectivities
and the transmittances of the ground and the absorber \cite{Then2008}.

\begin{figure}[t]
  \begin{minipage}{\columnwidth}
    \includegraphics[width=\columnwidth]{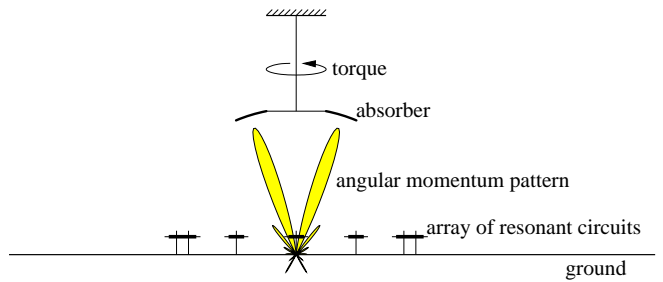}
  \end{minipage}
  \caption{%
    A modified version of the torque experiment that yields a larger
    torque on a smaller absorber.}
  \label{fig:4}
\end{figure}

There are two ways to facilitate the experiment.
Replacing the dipole antennas by resonant circuits with frequencies
in the kHz range increases the torque, because the angular momentum
is proportional to the inverse of the frequency.
Instead of using a half-sphere for the absorber,
a circular strip can be used that is smaller and that is
located where the angular momentum density has its maximum, see
Fig. \ref{fig:4}.

\section{Determining the angular momentum
  $\bfJ=\varepsilon_0\int d^3x(\bfx-\bfx_0)\times(\bfE\times\bfB)$
  with an array of tripole antennas}

Instead of measuring the torque directly, the electric and magnetic
field distributions can be measured with an array of tripole antennas
from which the angular momentum can be computed. Notice that if the
linear momentum $\bfP$ is non-zero, the angular momentum
$\bfJ=\varepsilon_0\int d^3x(\bfx-\bfx_0)\times(\bfE\times\bfB)$
depends on the reference point $\bfx_0$. The related quantity
$\grad\times(\bfE\times\bfB)$, or momentum vorticity, does not
have this problem. If vector $\bfE$ and $\bfB$ fields are measured
in an array, $\grad\times(\bfE\times\bfB)$ can be computed using
numerical discrete approximations of the differential.

Care has to be taken since the angular momentum is in the
non-transversal field components. In the far field, the transversal
components exceed the non-transversal components of the em field by
far and the latter become hidden in the noise.
However, even if the radial components become hidden, the angular
momentum is transported unaltered all the way to infinity
\cite{Abraham1914}. Hence, one needs other methods to
measure the POAM, see e.g. the rotational Doppler shift or the
Hanbury Brown and Twiss effect, further below.

Finally, we mention that the antenna array needs to cover a whole
cross section of the beam requiring huge arrays, if the distance
to the emitter becomes large. Consequently, better ways for detecting
the POAM are explored, below.

\section{Holograms}

In laboratory, holograms for radio waves are unfeasible; they are just
too large. Nevertheless, they may become important for experiments in
space. Just think of the spiral arms of a neighboring galaxy to form a
diffractive optics, or the possibility to design dust jets by special
flight patterns of rockets.

Holograms can replace antenna arrays. Instead of using an
array for triangulating the phases of the field, see below,
the beam vorticity can be converted by a hologram.
With a single crossed dipole antenna that is located on the beam axis
behind the hologram, it can be checked whether the
topological charge of the incoming beam is compensated by the hologram's
vorticity.

We are nevertheless aware of the fact that
holograms for em waves are experimentally challenging because of their large
size and the fact that they need to be aligned with the beam axis.

\section{Triangulation}

The POAM can be detected by triangulating the
electric field with an array of tripole antennas.
If the azimuthal phase shift associated with the topological charge
is known along a circle around the beam axis,
the POAM can be read off, cf. Fig. \ref{fig:2}. In the simplest setup this
requires that the beam axis is on the
antenna array and that the array extends out to the region where the doughnut
shaped beam carries its intensity.

In a pure POAM state one can read off the POAM from the azimuthal
phase shift directly. Any superposition of different POAM states is
decomposed via a discrete Fourier transform of the azimuthal phase shifts
\cite{Thide2007}.

\begin{figure}[t]
  \begin{minipage}{\columnwidth}
    \includegraphics[width=\columnwidth]{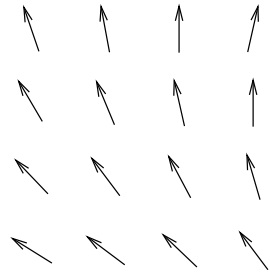}
  \end{minipage}
  \caption{%
    Schematic description of the instantaneous electric field vectors of an
    off axis radio beam with spin angular momentum $\bfS$ and orbital
    angular momentum $\bfL$, across a $4\times4$ antenna array.}
  \label{fig:5}
\end{figure}

There is no guarantee that the axis
of the incoming beam hits the antenna array. Typically, the
beam axis misses the array and the antennas cover only a part of the
doughnut shaped beam, see Fig. \ref{fig:5}.
Luckily, the phases of the em wave can be measured with high
accuracy allowing for their extrapolation. This enables off axis
measurements that are possible in radio only, but not in optics.

\begin{figure}[t]
  \begin{minipage}{\columnwidth}
    \includegraphics[width=\columnwidth]{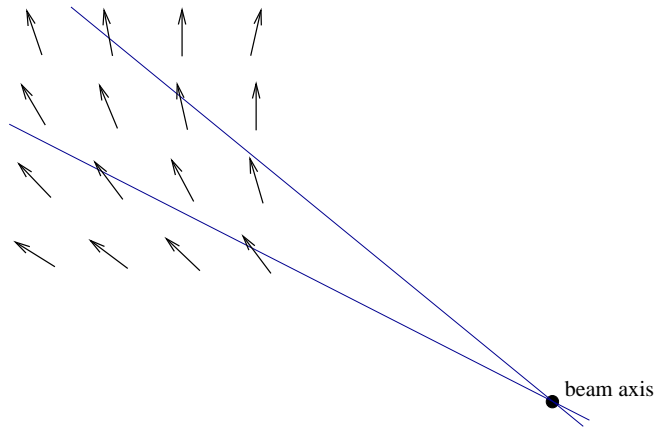}
  \end{minipage}
  \caption{%
    Same as Fig. \ref{fig:5}.
    In addition, lines of constant relative phases are shown.
    If extrapolated, the lines cross on the beam axis.}
  \label{fig:6}
\end{figure}

\begin{figure}[t]
  \begin{minipage}{\columnwidth}
    \includegraphics[width=\columnwidth]{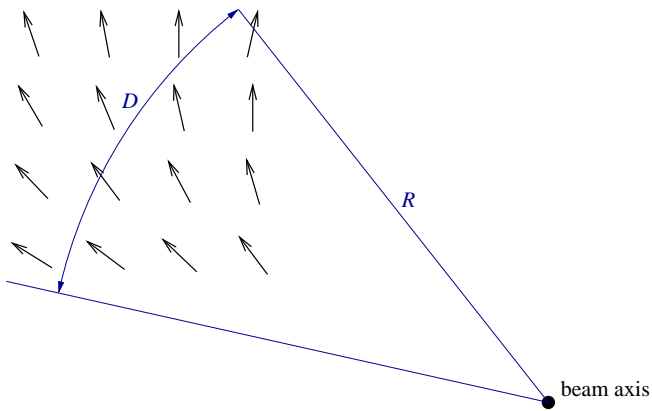}
  \end{minipage}
  \caption{%
    The circle segment that is covered by the antenna array.}
  \label{fig:7}
\end{figure}

For simplifying the explanation, it is assumed that the incoming
radio wave is in a pure POAM state.
If the axis of the incoming radio beam is not centered on
the antenna array, the extrapolation is carried out as follows:
(i)~One measures the relative phases of the $\bfE$ field at each antenna
of the finite size array, see Fig. \ref{fig:5}.
(ii)~One finds lines of constant relative phases by interpolating the
$\bfE$ field over the finite size array, see Fig. \ref{fig:6}.
(iii)~Extrapolating these lines far out until they cross in one point,
one finds the beam axis. The distance from the antenna array to the beam
axis is denoted by $R$.
(iv)~A circle segment with radius $R$ and angle $D/R$ is covered by the
antenna array, where $D$ is the diameter of the array, see Fig. \ref{fig:7}.
(v)~Neglecting all errors in the geometrical construction, except of the
uncertainty of the relative phase $\Delta\phi$ along the circle
segment of angle $D/R$, results in a $2\pi/(D/R)$ multiple thereof, if
extrapolated to the whole circle. This finally yields
$2\pi\Delta l>\Delta\phi2\pi/(D/R)$ or $\Delta l>R\Delta\phi/D$ as a
lower limit for the uncertainty in the extrapolated value of the
topological charge.
The error one is confronted with in a real observation is indeed larger,
because there is also the uncertainty in determining $R$.
Since the topological charge is integer valued, the uncertainty does
not matter as long as it is less than $1/2$. Off axis measurements allow
to estimate large values of the topological charge up to
$|l|<K\pi R/D$ with a moderate number $K$ of antennas along a circle
segment of length $D$.

Note that in case the emitter is at an astronomical distance,
i.e. extremely far away,
the diameter of the antenna array that needs to cover a substantial part
of the doughnut shaped beam profile is required to become giant,
see Fig. \ref{fig:8}.
An earth bound antenna array, if compared to the size of the
incoming radio signal, may be just like a local detector in one
point that is capable of measuring the spin part of the angular
momentum only.

The following example explains the situation more quantitatively.
Consider a source at the distance of Sirius, $r=2.637\,\text{pc}$,
that emits a radio signal at the wavelength of the hydrogen line,
$\lambda=21\,\text{cm}$, and that the emitting process shall take place
on a circular ring with a diameter that is comparable with the size of our
sun, $d=1.3914\cdot10^9\,\text{m}$, the diameter of the doughnut
shaped beam mode takes roughly the size of
$D=|l|r\frac{\lambda}{d}=11940\,\text{km}\ (l=\pm1)$ which is
comparable with the size of the earth.
In this case the POAM can be detected if the relative phases
of the incoming beam are triangulated across the whole surface of
the earth. But if the distance to the source increases, if the
wavelength becomes larger, or if the emitting process is in a smaller
region, the limits of what is detectable with current techniques of
triangulation are exceeded.

\section{The lighthouse effect}

If the incoming em waves are coherent and stationary, one can use
a single tripole antenna to explore the electric field in space by moving
the antenna around and recording the data. This allows for
triangulation methods while using one single antenna only.

\begin{figure}[t]
  \begin{minipage}{\columnwidth}
    \includegraphics[width=\columnwidth]{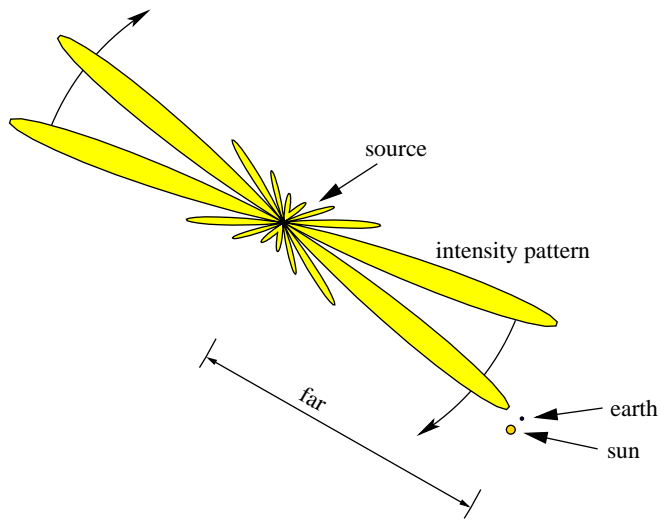}
  \end{minipage}
  \caption{%
    The doughnut shaped radiation pattern of a rotating source traversing
    an earth-bound telescope.}
  \label{fig:8}
\end{figure}

If the emitter rotates around an axis that differs from the beam axis,
one does not need to move the antenna
around, because the em fields themselves move across the detector.
Let us call this effect the light house effect.
Mapping the spatial azimuthal phase correlations to
temporal phase correlations allows to detect the POAM of an arbitrary far
distant source with a single dish telescope, see Fig. \ref{fig:8}.

The key advantage of the light house effect is that it increases the
``POAM baseline'' of the telescope to an enormous size, depending on
the coherence time but not on the physical size of the telescope.

\section{Radiation patterns}

From the radiation pattern, i.e. the distribution of energy which can
be measured in the far field, one can conclude via models
of the emitting process whether the
beam carries POAM. For instance, a doughnut shaped
radiation pattern gives a strong hint that the topological charge
should be non-zero.
Since a model for the radiation process has to be assumed, the
conclusion of POAM from radiation patterns is highly indirect
and may also not be unique.

\section{Rotational Doppler effect}

Should the radio emitting object rotate fast and have sharp
discrete lines in its emission spectrum, the angular momentum of the
emitted em waves can be measured indirectly via shifts and splittings of
the spectral lines. The shifts result from the rotational Doppler
effect $\omega-\omega_0=(l+s)\Omega_\parallel$ where $\Omega_\parallel$ is the
projection of the rotation frequency onto the wave vector $\bfk$
\cite{Courtial1998}. Decomposing into pure spin states, the
discrete emission spectrum will be decomposed into one for $s=+1$ and
one for $s=-1$.  These two spectra should almost coincide in their
spectral lines, except for an overall shift.  The relative overall shift
between the two spectra is equal to twice the rotational frequency of
the emitter.  Once $\Omega_\parallel$ has been read off, one can search
for spectral lines that are separated exactly by $\Omega_\parallel$ and
integer multiples thereof.  Each of these spectral lines corresponds to
a specific POAM state \cite{Thide2007}.

\section{The Hanbury Brown and Twiss effect}

Higher order quantum correlations in the temporal domain
allow for intensity interferometry of light and em waves,
as was found and demonstrated by Hanbury Brown and Twiss.
The original derivation of Hanbury Brown and Twiss for their effect
is valid, if the incoming radiation is in a pure quantum state.
Different polarization states (spin) give an extra factor of one-half,
see Hanbury Brown and Twiss \cite{Hanbury1957a}, page 319, lines 7--17.
That the em waves can also differ in state, because of POAM,
has been overseen. The latter gives raise to
another factor which depends on the geometry of the doughnut shaped
beam, namely on how many doughnuts of individual POAM states overlap
with each other at the detector.

\section{Summary}

Methods for the detection of POAM in the radio domain have now been
proposed. Our list is not exhaustive and we understand it rather as
a door opener to an unexplored area of fundamental physics. While
some experiments are suitable to be conducted in a laboratory, e.g.
the direct torque measurement, the triangulation of the phases of
the em field requires large detectors.
If the progress of the upcoming radio telescopes LOFAR \cite{lofar},
LOIS \cite{lois}, and SKA \cite{ska} keeps on going, these telescopes
will become in the future the natural candidates for systematically
scanning the sky for POAM in radio via triangulating the electric field.

\section*{Acknowledgments}

The authors thank Martin Harwit for useful discussions.
In particular, it was him who first mentioned the Hanbury Brown and Twiss
effect in connection with POAM.
Part of the work was supported by the Centre for Dynamical Processes and
Structure Formation, Uppsala University, Sweden.

\end{document}